\documentclass[sigplan,nonacm]{acmart}

\makeatletter
\def\city#1{\global\@ACM@citypresenttrue}
\def\country#1{\global\@ACM@countrypresenttrue}
\makeatother

\newcommand{\thiswork}{\textsf{TLX}}

\newcommand{\Sec}[1]{Sec.~{#1}}
\newcommand{\Fig}[1]{Fig.~{#1}}
\newcommand{\Tbl}[1]{Tbl.~{#1}}

\usepackage{graphicx} 
\usepackage{amsmath}

\usepackage{lipsum}  

\usepackage{makecell}

\usepackage{booktabs}

\usepackage{array}

\usepackage{makecell}

\usepackage[table,xcdraw]{xcolor}

\usepackage{enumitem}

\usepackage{circledtext}

\usepackage{multirow}

\usepackage[most]{tcolorbox}
\definecolor{LightBlue}{RGB}{218,227,243}
\definecolor{VeryLightBlue}{RGB}{238,241,255}
\definecolor{VeryLightGray}{gray}{0.95}
\definecolor{CodeKeyword}{RGB}{0,70,140}
\definecolor{CodeComment}{RGB}{0,120,70}
\definecolor{CodeString}{RGB}{163,21,21}
\usepackage{listings}

\newtcblisting[auto counter]{CodeListing}[3][]{
enhanced,
listing engine=listings,
listing only,
listing options = {
  language=Python,
    #3
},
overlay={%
    \begin{tcbclipinterior}
        \fill[gray!25] (frame.south west) rectangle ([xshift=4mm]frame.north west);
    \end{tcbclipinterior}
},
frame hidden,
fonttitle=\fontfamily{\sfdefault}\selectfont,
colback=VeryLightGray,left=4mm,
colbacktitle=LightBlue,coltitle=black,
title={Listing \thetcbcounter: #1},
#2
}
\newtcolorbox{CodeListingInSubfigure}[1][]{
enhanced,breakable,
fonttitle=\fontfamily{\sfdefault}\selectfont,
left=1pt,right=0pt,top=0pt,bottom=0pt,
#1
}

\begin{document}
\title{TLX: Hardware-Native, Evolvable MIMW GPU Compiler for Large-scale Production Environments}

\author{Yue Guan}
\affiliation{%
	\institution{UC San Diego}
	\city{La Jolla}
	\country{USA}
}
\email{y9guan@ucsd.edu}

\author{Hongtao Yu}
\affiliation{%
	\institution{Meta}
	\city{Menlo Park}
	\country{USA}
}
\email{hoy@meta.com}

\author{Peng Chen}
\affiliation{%
	\institution{Meta}
	\city{Menlo Park}
	\country{USA}
}
\email{pchen7e4@meta.com}

\author{Daohang Shi}
\affiliation{%
	\institution{Meta}
	\city{Menlo Park}
	\country{USA}
}
\email{daohang@meta.com}

\author{Karthik Manivannan}
\affiliation{%
	\institution{Meta}
	\city{Menlo Park}
	\country{USA}
}
\email{kmanivannan@meta.com}

\author{Nicholas J Riasanovsky}
\affiliation{%
	\institution{Meta}
	\city{Menlo Park}
	\country{USA}
}
\email{njriasan@meta.com}

\author{Manman Ren}
\affiliation{%
	\institution{Meta}
	\city{Menlo Park}
	\country{USA}
}
\email{mren@meta.com}

\author{Lei Wang}
\affiliation{%
	\institution{Meta}
	\city{Menlo Park}
	\country{USA}
}
\email{wlei@meta.com}

\author{Shane Nay}
\affiliation{%
	\institution{Meta}
	\city{Menlo Park}
	\country{USA}
}
\email{snay@meta.com}

\author{Partha Kanuparthy}
\affiliation{%
	\institution{Meta}
	\city{Menlo Park}
	\country{USA}
}
\email{pka@meta.com}

\author{Zhijing Li}
\affiliation{%
	\institution{Meta}
	\city{Menlo Park}
	\country{USA}
}
\email{tissue030@meta.com}

\author{Ying Liu}
\affiliation{%
	\institution{Meta}
	\city{Menlo Park}
	\country{USA}
}
\email{liuying@meta.com}

\author{Zaifeng Pan}
\affiliation{%
	\institution{UC San Diego}
	\city{La Jolla}
	\country{USA}
}
\email{zapan@ucsd.edu}

\author{Zhengding Hu}
\affiliation{%
	\institution{UC San Diego}
	\city{La Jolla}
	\country{USA}
}
\email{zhh068@ucsd.edu}

\author{Yufei Ding}
\affiliation{%
	\institution{UC San Diego, Meta}
	\city{La Jolla}
	\country{USA}
}

\email{yufeiding@ucsd.edu}
\renewcommand{\shortauthors}{Y. Guan et al.}

\begin{abstract}
Modern GPUs increasingly rely on specialized hardware units and asynchronous coordination mechanisms, so performance depends on orchestrating data movement, tensor-core computation, and synchronization rather than exposing more thread-level parallelism. This creates a programming-model tension: if too much execution structure is hidden, the compiler must catch up to new hardware mechanisms; if too much is exposed, the burden of orchestration falls back onto the programmer. We present \emph{\thiswork{}} (Triton Low-level Language Extensions), built around \emph{MIMW} (\emph{Multi-Instruction, Multi-Warp}), which expresses orchestration at warp-group granularity while preserving Triton's productive blocked programming model for regular computation. TLX realizes this idea as an embedded extension to Triton, exposing explicit interfaces for multi-warp execution, local-memory orchestration, asynchronous operations, and cluster-aware control. Our evaluation shows that TLX supports substantial customization with limited development effort while remaining competitive with state-of-the-art implementations. TLX-authored kernels have been deployed in large-scale training and inference production systems. Our code is open sourced at \url{https://github.com/facebookexperimental/triton}.
\end{abstract}

\maketitle 

\section{Introduction}
\label{sec:introduction}

Modern GPU performance is increasingly determined by how quickly programming systems evolve with hardware~\cite{tillet2019triton,cutlass}. Across recent GPU generations, two trends have become especially important: increasing \emph{specialization} and increasing \emph{asynchrony}. GPUs now contain more components that serve distinct roles, including data-movement engines~\cite{nvidia_hopper_architecture}, tensor-memory pathways~\cite{nvidia_blackwell_architecture}, and specialized matrix units~\cite{nvidia_ampere_a100_whitepaper}. To keep these components fully utilized, hardware increasingly provides asynchronous mechanisms that let them run and communicate without forcing the whole processor into one lockstep execution pattern. The programming challenge therefore shifts from simply exposing more thread-level parallelism to coordinating specialized components, their data movement, and dependences efficiently.

\begin{figure*}
\centering
    \includegraphics[width=\linewidth]{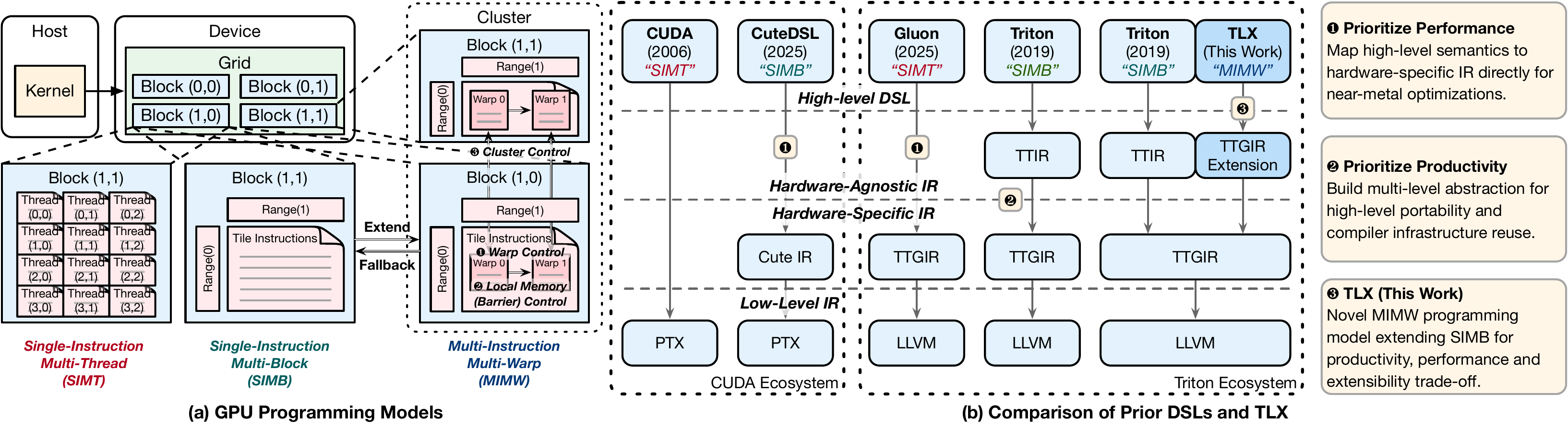}
    \vspace{-2em}
    \caption{MIMW as the missing middle between SIMT and SIMB, and TLX's two-layer realization within the Triton ecosystem.}
    \label{fig:tlx_overview}
\end{figure*}

The first widely adopted answer was the \emph{single-instruction, multi-thread} (\emph{SIMT}) programming model~\cite{SPMD}, exemplified by CUDA~\cite{nvidia_cuda_programming_guide}. In this model, programmers directly manage threads, synchronization, memory movement, and hardware-specific execution structure. Because control is exposed at a very fine granularity, SIMT can map closely to the machine and deliver excellent performance without waiting for a compiler to infer every orchestration decision. But it does so by pushing the full burden of orchestration onto the programmer and by giving up much of the leverage provided by higher-level blocked abstractions and compiler automation.

The next major step was to raise the abstraction level from threads to blocked programs through a \emph{single-instruction, multi-block} (\emph{SIMB}) style~\cite{triton}. Systems such as Triton were developed to address the productivity limitations of CUDA by hiding intra-block orchestration behind a blocked abstraction while preserving high performance on regular kernels. More recently, NVIDIA has also pushed this direction within the CUDA ecosystem through CuTeDSL~\cite{cute_dsl}. This design is effective when a compiler can recover the required low-level schedule, but that same abstraction boundary also means that when hardware introduces new intra-SM execution patterns, asynchronous mechanisms, or coordination scopes, the compiler must discover and implement all of them on the user's behalf. In practice, many of these orchestration decisions are not statically inferable within reasonable compilation cost. As a result, reaching state-of-the-art performance can require a long compiler catch-up cycle even when the necessary hardware optimizations are conceptually clear\cite{chen2026tawa, soi2025optimal}.

As GPU hardware has continued to evolve, this SIMB direction has exposed a new bottleneck: compiler catch-up. If a programming model hides too much, then every new orchestration pattern must be rediscovered by compiler analysis and lowering; if it exposes too much, then the full burden of scheduling and synchronization falls back onto the programmer. The recurring Renaissance of lower-level systems such as Gluon\cite{triton_gluon} reflects this pressure: when Triton-style abstractions cannot expose enough control quickly enough, developers seek more explicit interfaces to recover performance. \emph{The central programming-model question is therefore how to divide responsibility between the programmer and the compiler in a way that can keep pace with hardware evolution.}

TLX is our answer to this next stage of DSL development. It is built around \emph{MIMW} (\emph{Multi-Instruction, Multi-Warp}), a middle position in which orchestration is expressed at warp-group granularity rather than purely at block or thread granularity. We observe that modern GPU kernels are neither naturally block-uniform nor naturally thread-by-thread. The critical control point is the \emph{warp group}: coarse enough that each group can own a specialized hardware role, but fine enough to express asynchronous producer--consumer pipelines inside a streaming multiprocessor (SM)~\cite{lindholm2008tesla}. In MIMW, different warp groups execute distinct instruction streams while cooperatively orchestrating data movement and synchronization. As illustrated in Figure~\ref{fig:tlx_overview}, MIMW sits between SIMB and SIMT: it keeps blocked computation in a form that compilers handle well, while exposing the orchestration decisions that programmers most need to control.

TLX realizes this idea as an extension to Triton rather than a replacement for it. Programmers can keep Triton's blocked programming style for the regular parts of a kernel and introduce explicit multi-warp orchestration only where the hardware requires it. This design preserves the existing compiler stack where it is already effective, provides a natural fallback to plain Triton when explicit control is unnecessary, and offers a path we believe is more sustainable for future GPU DSLs than either returning fully to CUDA-style programming or relying on compiler catch-up alone.

This combination of blocked computation and explicit orchestration is what lets TLX target both productivity and performance. The upper layer preserves Triton's productive blocked model, while the extension layer makes asynchronous pipelines, local-memory orchestration, and hardware-specific execution structure natural, first-class extensions of that same programming model. In this way, TLX gives programmers a unified path from regular blocked kernels to hardware-tailored execution, reducing development effort without forcing performance-critical kernels to wait for a full compiler rediscovery of each new optimization opportunity.

We evaluate TLX in large-scale ML production environments that motivate its design. The evaluation goes beyond canonical operator benchmarks~\cite{goto2008anatomy,shah2024flashattention,zadouri2026flashattention} to include production-oriented kernels with heterogeneous orchestration patterns~\cite{chang2024flux, roy2025fast}, where fast development and production-level performance are both required. The results show that TLX can support substantial customization with limited development effort while remaining competitive with state-of-the-art implementations. The rest of the paper develops the MIMW model and its primitives, describes TLX's integration into the Triton ecosystem, and presents this evaluation in detail.

\begin{figure*}
\centering
    \includegraphics[width=\linewidth]{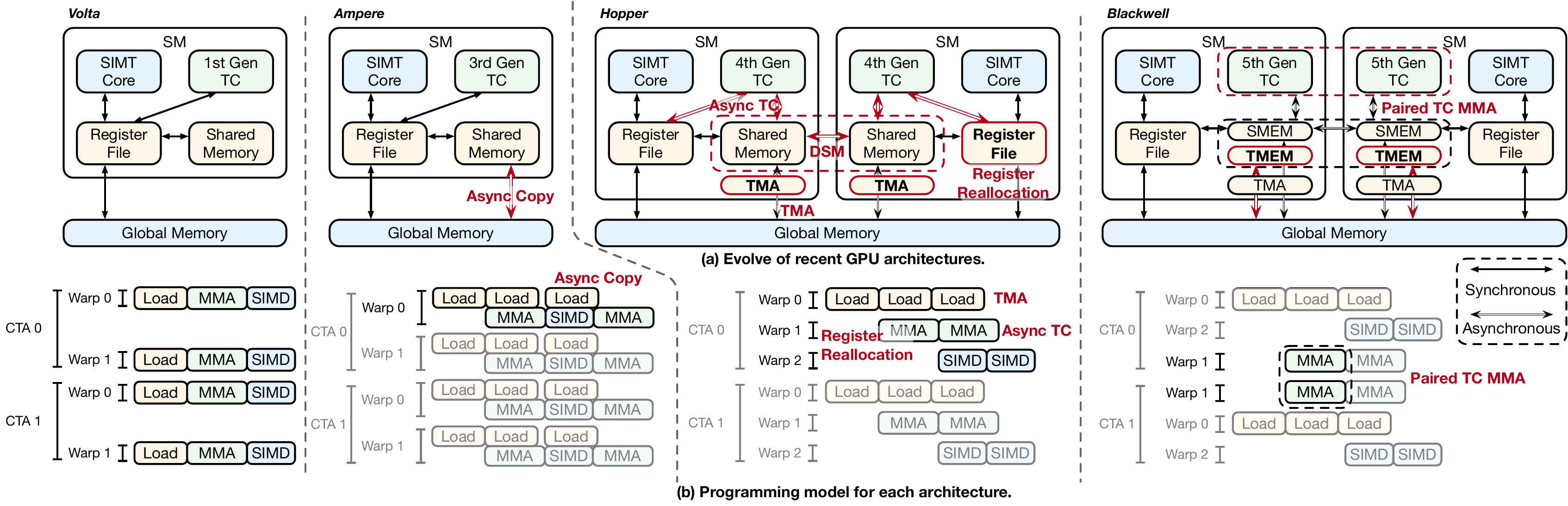}
    \vspace{-2em}
    \caption{GPU architecture and execution model.}
    \label{fig:gpu_arch}
\end{figure*}

\vspace{0.5em}
To summarize, the paper contributes in three main ways:

\begin{itemize}[leftmargin=*]
\vspace{-0.5em}
    \item We propose \emph{MIMW}, a programming model that identifies warp-group execution with explicit inter-group dependences as the missing middle ground between productivity-oriented SIMB abstractions and control-oriented SIMT programming.
    \item We design and implement \thiswork{} as a two-layer embedded extension to Triton that realizes MIMW in practice, exposing warp-group specialization, asynchronous operations, local-memory orchestration, and fine-grained synchronization while preserving Triton's productive blocked programming style when explicit control is unnecessary.
    \item We benchmark \thiswork{} on canonical and production kernels, showing CUDA-competitive performance with a more productive and extensible programming model.
\end{itemize}

\section{Background and Related Work}
This section summarizes the GPU architecture and programming model background.

\begin{table}[t]
\caption{Summary of GPU programming terminology}
\vspace{-1em}
\label{tbl:terminology}
\resizebox{\linewidth}{!}{
\begin{tabular}{lll}
\Xhline{3\arrayrulewidth}
Nvidia                        & AMD               & Description                                    \\ \hline
\rowcolor[HTML]{EFEFEF} 
\begin{tabular}[c]{@{}l@{}}Streaming \\Multiprocessor (SM)\end{tabular}  &  
\begin{tabular}[c]{@{}l@{}}Compute \\Unit (CU)\end{tabular}  & Parallel vector processors in GPUs.             \\
Thread                        & Work Item         & Individual lanes in a kernel.                     \\
\rowcolor[HTML]{EFEFEF} 
Warp                          & Wavefront         & A vector of threads run the same instructions. \\
Warp Group                         & -         & A group of collaborative warps. \\
\rowcolor[HTML]{EFEFEF} 
-                         & XCD Group         & A group of waves in the same chiplet. \\
Thread Block                  & Workgroup         & A group of warps executed at the same time.      \\ \Xhline{3\arrayrulewidth}
\end{tabular}
}
\end{table}

\subsection{GPU Architecture}

Although the proposed methodologies support both Nvidia and AMD GPUs, we use Nvidia terminology unless AMD-specific features~\cite{amd2025cdna4} matter. 
Modern GPUs are throughput-oriented accelerators with many parallel execution units and a deep memory hierarchy. Kernels launch as grids of thread blocks, or cooperative thread arrays (CTAs), on Nvidia and workgroups on AMD, scheduled onto streaming multiprocessors (SMs) or compute units (CUs). Within each block, threads execute in lockstep groups---warps on Nvidia and wavefronts on AMD---under the classic SIMT model (\Tbl{\ref{tbl:terminology}}). Each SM/CU contains SIMT cores, a large register file, and software-managed on-chip memory (shared memory / LDS) for data reuse and inter-thread cooperation.

\Fig{\ref{fig:gpu_arch}} shows that recent GPU generations are moving beyond simple SIMT toward more \emph{specialized}, \emph{asynchronous}, and \emph{collaborative} execution. Architectures add dedicated units beyond general SIMT cores, including tensor cores, asynchronous copy engines, and tensor memory accelerators (TMA); these units let data movement, tensor computation, and SIMT execution overlap as pipeline stages; and execution is shifting from mostly independent warps to coordinated warp groups that specialize in different tasks and synchronize explicitly.

This shift also appears in tensor-core instructions. As summarized in \Tbl{\ref{tbl:tensor_core_evolution}}, the programming granularity expands from Volta's quad-level MMA to Ampere's warp-level WMMA, Hopper's warp-group WGMMA, and Blackwell's CTA-pair tcgen05.mma~\cite{luo2025dissecting}. Newer tensor-core pipelines require cooperation from increasingly larger thread groups.

Together, these changes make the original SIMT mental model increasingly insufficient for performance reasoning. High performance now depends not only on parallelizing work, but also on orchestrating specialized units, overlapping asynchronous stages, and coordinating execution across warps or warp groups. This trend motivates programming models and compiler support that expose structured control over asynchronous execution and cross-warp cooperation.

\begin{table}[b]
\caption{Evolution of Nvidia Tensor Core Instructions}
\vspace{-1em}
\label{tbl:tensor_core_evolution}
\resizebox{\linewidth}{!}{
\begin{tabular}{llll}
\toprule
Architecture & Instruction & Size(MNK)   & Granularity              \\
\midrule
\rowcolor[HTML]{EFEFEF} 
Volta        & MMA        & 8-8-4       & Quapair (8 threads)      \\
Ampere       & WMMA        & 16-8-16     & Warp (32 threads)        \\
\rowcolor[HTML]{EFEFEF} 
Hopper       & WGMMA       & 64-8-16     & Warp Group (128 threads) \\
Blackwell    & tcgen05.mma & 256-256-128 & CTA Pair (256 threads) \\
\bottomrule
\end{tabular}
}
\end{table}

\subsection{GPU Programming Models and DSLs}

\begin{figure*}
    \centering
    \includegraphics[width=\textwidth]{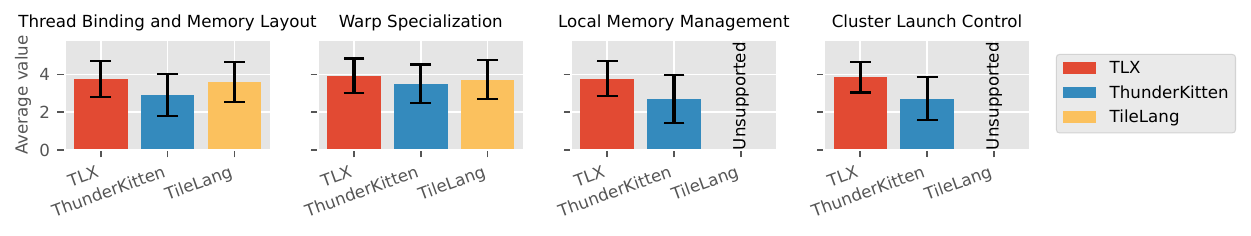}
    \vspace{-3em}
    \caption{Productivity survey results.}
    \label{fig:productivity-survey}
\end{figure*}

This subsection focuses on one dimension of GPU DSL design: the granularity at which execution structure is represented. Recent hardware trends make this choice increasingly important, because high-performance kernels must now coordinate specialized units, asynchronous stages, and cooperative groups that are larger than a single warp but more structured than an entire block.

At one end of the spectrum is the traditional \emph{SIMT} model, represented on the SIMT side of \Fig{\ref{fig:tlx_overview}}. CUDA\cite{nvidia_cuda_programming_guide} exposes this model directly: programmers write threads organized into blocks, and warps execute a largely uniform instruction stream over different data elements. Its main strength is explicit control over the execution hierarchy, memory spaces, and synchronization primitives. That control makes SIMT fully expressive for modern kernels, but it also means that asynchronous copies, tensor-core pipelines, staging buffers, and cross-warp coordination must be orchestrated explicitly by the programmer.

At the other end is Triton's \emph{SIMB} style, represented on the SIMB side of \Fig{\ref{fig:tlx_overview}}. Triton\cite{triton} raises abstraction from threads to blocked tensor programs: the programmer writes one tile-centric program, and the compiler replicates and optimizes it across blocks. This improves productivity substantially and works well for many dense ML operators. It establishes a different contract between user and compiler: the user specifies what a block computes, while the compiler decides how that computation maps to the machine. As execution becomes more asynchronous and role-specialized, however, this abstraction makes it harder to represent kernels whose performance depends on distinct roles within a block, explicit asynchronous dependencies, or coordination patterns that don't fit one uniform block-level instruction stream.

Recent concurrent systems such as ThunderKitten\cite{spector2024thunderkittens} and TileLang\cite{wang2025tilelang} push in a related direction by extending the SIMB-style interface with more low-level control over kernel structure and scheduling. They show that the community is converging on the need for more explicit orchestration within tile-based DSLs. Their design point is nevertheless different from TLX's: they are not extensions of the Triton ecosystem itself, and they do not expose the same set of low-level features that we study in our productivity survey, especially around cluster launch control and other explicit orchestration mechanisms. In that survey, 127 graduate-level students in a GPU programming course compared TLX against other GPU programming systems along four dimensions that directly reflect the controls discussed here: thread binding and memory layout, warp specialization, local memory management, and cluster launch control; the detailed survey setup is provided in \Sec{\ref{sec:appendix:productivity_survey}}. As shown in \Fig{\ref{fig:productivity-survey}}, TLX remains competitive on the common low-level tasks and stands out most clearly on cluster launch control, suggesting that the additional orchestration controls exposed by TLX remain usable in practice rather than forcing programmers back to fully manual SIMT-style coding.



\section{\thiswork{} Overview} \label{sec:overview}

This section presents TLX's design at a high level. The goal is not to reintroduce every mechanism in detail, but to make the organizing idea clear: TLX keeps Triton's productive blocked model for tile computation while exposing the execution and memory structure that modern kernels increasingly need to control explicitly. A small end-to-end example then shows how these abstractions compose at the source level.

\subsection{Embedded Language Design}


\Fig{\ref{fig:tlx_overview}} summarizes TLX as an additive extension to Triton rather than a replacement language. The \emph{Triton layer} keeps Triton's tile-based frontend and optimization pipeline. The \emph{TLX extension layer} exposes execution and memory structure that Triton's default SIMB abstraction intentionally hides: warp specialization, asynchronous dependencies, local-memory orchestration, and cluster-aware coordination.

The key idea is simple. Triton gives programmers a productive way to describe what tile of work a kernel computes; TLX adds a way to describe how cooperating warp groups and CTAs stage, synchronize, and coordinate that work when performance depends on explicit orchestration. The important point is that this interface is a realization of MIMW, not the definition of MIMW itself. MIMW is the abstraction boundary that makes warp-group roles, cross-role dependences, and local-state orchestration first-class; TLX is the embedded language and compiler path that carries those constructs through Triton's stack. In practice, this means that concurrency becomes a source-level concept, dependencies become explicit rather than implicit in lockstep execution, and on-chip storage becomes a compiler-visible object with well-defined lifetime and layout.

\begin{CodeListing}[\thiswork{} overview example.]{label=lst:interface,float}{basicstyle=\scriptsize\ttfamily}
@triton.jit
def tlx_kernel(x_ptr, y_ptr, n_elements, BLOCK: tl.constexpr):
    tile_id = tl.program_id(0)
    full = tlx.alloc_barrier(1, arrive_count=1)
    empty = tlx.alloc_barrier(1, arrive_count=1)    

    # Sec.4.2 Local Memory Management
    smem = tlx.local_alloc((BLOCK,), tl.float32, 1)

    # Sec.4.3 Cluster Control
    clc = tlx.clc_create_context(num_consumers=4)
    cta_rank = tlx.cluster_cta_rank()

    # Sec.4.1 Warp Specialization
    with tlx.async_tasks():
        with tlx.async_task("default"):
            while tile_id != -1:
                tlx.barrier_wait(empty[0], prod_phase)
                x = tl.load(x_ptr+...)
                tlx.local_store(smem[0], x)
                tlx.barrier_arrive(full[0])
                tlx.clc_producer(...)
                tile_id = tlx.clc_consumer(...)
        with tlx.async_task(num_warps=1):
            while tile_id != -1:
                tlx.barrier_wait(full[0], cons_phase)
                x_local = tlx.local_load(smem[0])
                y = x_local * 2.0 + cta_rank
                tl.store(y_ptr+..., y)
                tlx.barrier_arrive(empty[0])
                tile_id = tlx.clc_consumer(...)

# Kernel Launch
tlx_kernel[grid](x, y, n, 
    BLOCK=256, num_warps=4, ctas_per_cga=(2, 1, 1),)
\end{CodeListing}

\begin{figure*}
    \centering
    \includegraphics[width=\linewidth]{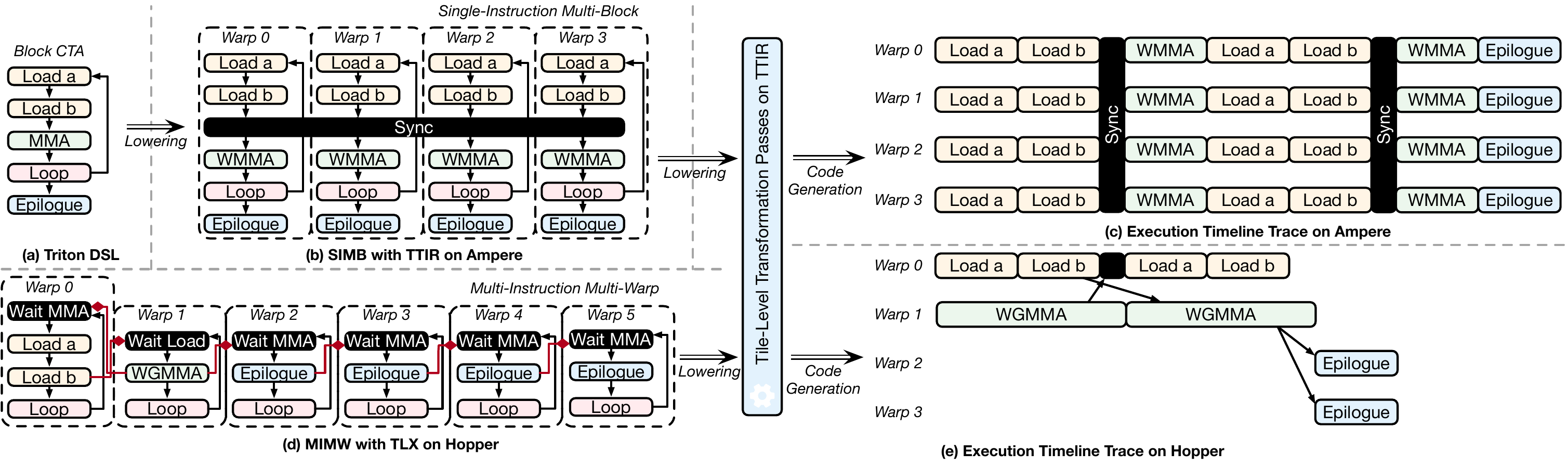}
    \vspace{-2.5em}
    \caption{Warp specialization and its integration with the MIMW programming model.}
    \label{fig:ws}
\end{figure*}

This organization lets TLX pursue the three goals from the introduction without falling back to opaque backend intrinsics. Productivity comes from leaving tile computation in Triton's existing model. Performance comes from making coordination and staging explicit where needed. Extensibility comes from representing those decisions in the frontend and IR, so new hardware mechanisms can be lowered systematically instead of through ad hoc escape hatches. That systems story is separate from the conceptual claim: the paper argues that MIMW is the right abstraction boundary, and TLX demonstrates that the boundary can be embedded cleanly into an existing blocked DSL rather than requiring a separate low-level language.

\subsection{Embedded Language Interface}

Listing~\ref{lst:interface} shows a minimal TLX kernel that exercises this design. One task stages data into local memory, another consumes it, the two tasks synchronize through mbarriers, and cluster control provides work assignment. The example is intentionally small; its purpose is not to model one full application kernel, but to show the shape of the source-level interface when execution structure must be made explicit.

This example is best read as one small pipeline rather than a checklist of features. The \texttt{tlx.async\_tasks()} region says that a CTA contains multiple concurrent roles. The producer role stages data into an explicit local buffer, the consumer role waits only on the dependency it needs, and the cluster intrinsics show that the same interface can express dynamic work assignment beyond a single CTA. In other words, the listing illustrates TLX's main claim: tile computation stays in Triton's familiar style, while scheduling, synchronization, and staging become explicit where needed.

The point of the example is therefore not to define every primitive exhaustively; Section~4 does that. Its job here is to make the language boundary visible. Programmers still write a high-level kernel, but they can now surface warp roles, synchronization edges, local staging, and cluster coordination directly in the program when the hardware and algorithm require it.


\begin{figure*}
    \centering
        \includegraphics[width=\linewidth]{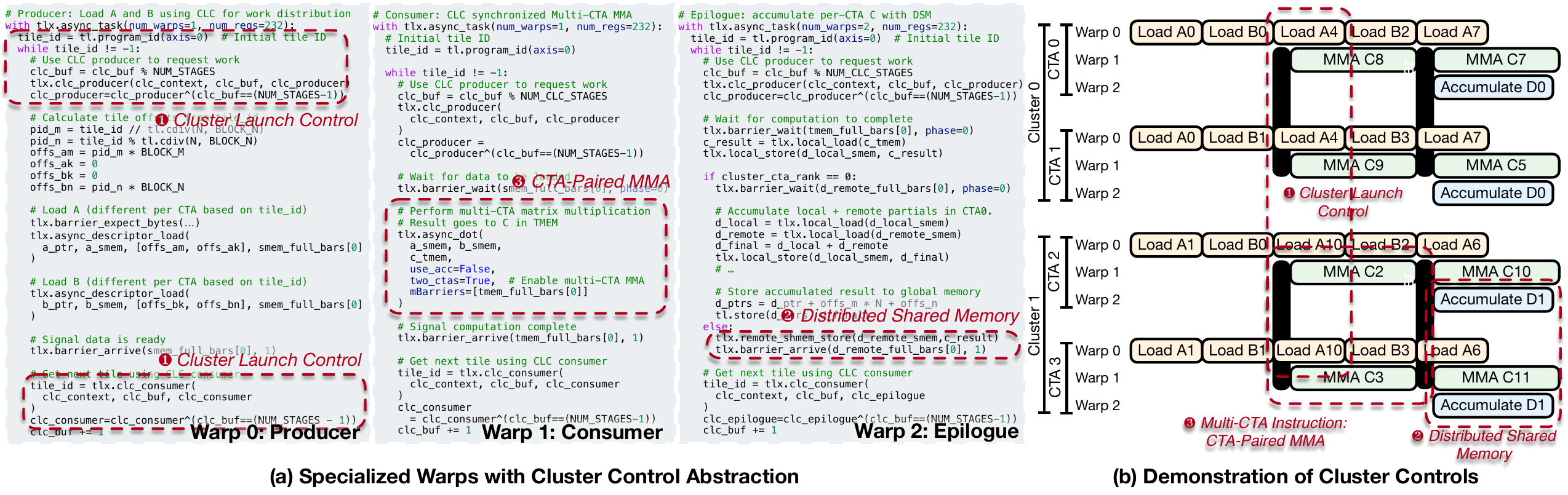}
        \vspace{-2em}
    \caption{Cluster-level control in TLX. The left illustrates source-level interfaces, while the right shows how these mechanisms are driven by specialized warp roles within the MIMW execution model.}
    \label{fig:cluster-control}
\end{figure*}
\section{MIMW with TLX}

This section describes how TLX realizes MIMW during lowering. The three mechanisms form a layered control contract: warp-level control makes execution roles explicit within a CTA, cluster-level control lifts role-specialized tasks across CTAs, and local-memory control makes intermediate state, layout constraints, and synchronization edges explicit to the compiler. Because each layer is represented directly in the source model and IR rather than buried in backend code patterns, the same structure can be retargeted as future hardware introduces new scheduling scopes, communication primitives, or local-memory behaviors.

\subsection{Warp-Level Control}
\label{subsec:ws}

Warp specialization~\cite{10.1145/2063384.2063400,crago_wasp_2024,bauer_singe_2014} maps async tasks onto disjoint warp groups within one CTA. It is the base control layer in TLX: once warp groups and their dependencies are explicit, later mechanisms can extend the same structure outward to CTA clusters and downward to the local state those tasks exchange. Instead of cloning one block program across all warps, TLX lets the programmer define role-specialized regions such as producer, consumer, MMA, and epilogue paths and connect them with explicit dependencies.

\vspace{0.5em}
\noindent \textbf{Integration with MIMW.}
\Fig{\ref{fig:ws}} shows why this execution style needs a representation beyond Triton's default lowering. In standard Triton, the programmer writes one tile-level program, and the compiler effectively replicates it across warps within the CTA. This \emph{single-instruction multi-warp} organization is highly productive when all warps should execute the same sequence. It does not directly express the case where different warp groups intentionally follow different instruction streams so that data movement, compute, and epilogue can overlap in time.

TLX adds warp specialization as an extension to Triton rather than a replacement. At the Python level, a TLX kernel remains a standard \texttt{@triton.jit} function, but the programmer can partition the CTA's warps into \emph{warp groups} and attach different regions to them. Conceptually, the kernel becomes a collection of cooperating warp-level tasks: a producer region issues asynchronous loads and signals readiness, a consumer region waits on those signals and performs MMA, and other regions may wait on compute completion and execute the epilogue. As the bottom row of \Fig{\ref{fig:ws}} shows, warps are no longer symmetric replicas; they are role-specialized participants connected by explicit \texttt{wait}/\texttt{arrive} dependencies.

Crucially, TLX keeps Triton's tile-centric compiler model. Each MIMW region is still written with Triton's tensor and pointer abstractions for the tile computations it performs, so most of Triton's TTIR/TTGIR optimization pipeline remains applicable within each region. During lowering, TLX generalizes Triton's original warp-partitioning passes: instead of cloning one instruction stream to all warps, it materializes multiple per-warp-group instruction streams, preserves the explicit cross-warp dependencies, and then passes the resulting TTGIR to the same downstream tile-level optimizations.

\subsection{Cluster-Level Control}
\label{subsec:cluster_control}
Warp specialization defines how roles are organized \emph{within} a CTA. Cluster-level control extends that structure across CTAs. CUDA thread-block clustering provides the execution domain, and TLX exposes scheduling, communication, and collective mechanisms needed to program it explicitly. In this way, cluster control is not a separate programming model layered beside warp specialization; it is the cluster-scoped continuation of the same role-specialized execution model.

\vspace{0.5em}
\noindent \textbf{Cluster abstraction in MIMW.}
\Fig{\ref{fig:cluster-control}} summarizes how TLX lifts MIMW execution from one CTA to a cooperating CTA cluster. Each CTA is treated as an independent tile, and clustered behavior is introduced through explicit inter-CTA communication and synchronization rather than inferred from layout metadata. Data sharing, work redistribution, and ordering constraints are therefore visible in the program structure. This formulation also supports CTA specialization: not all CTAs are required to participate in the same tile computation, and different CTAs within a cluster can assume distinct roles, each of which can in turn be implemented by specialized warp groups.

Cluster Launch Control (CLC), distributed shared memory (DSM), and multicast are all asynchronous control and communication mechanisms. Queue request and response in CLC proceed independently of computation; multicast TMA transfers and their barrier signaling can overlap with MMA; and DSM producers and consumers must coordinate ordering without forcing CTA-wide stalls. Expressing these overlaps is difficult in a single-instruction multi-warp model in which all warps execute the same loop body and synchronize CTA-wide at phase boundaries. TLX's MIMW model makes cluster control compositional by allowing one warp group to specialize in \emph{cluster progression}---advancing the CLC pipeline, issuing multicast TMA, or performing remote DSM stores and arrivals---while other warp groups continue executing data movement and compute for the current tile. Dependencies are expressed with explicit \texttt{wait}/\texttt{arrive} edges between warp-specialized regions, so only the warps that truly depend on a given event stall.

\vspace{0.5em}
\noindent \textbf{Cluster Launch Control.}
Clustering changes how work should be scheduled. When a cluster becomes resident, repeatedly launching short-lived kernels, or statically assigning each CTA a fixed tile, can leave resources underutilized because of irregular tile runtimes, tail effects, or data-dependent control flow. CLC addresses this problem by enabling \emph{dynamic persistent} execution: CTAs or clusters repeatedly acquire new tile identifiers from a hardware-managed work queue, providing implicit work stealing and load balancing without a software queue or global coordination. TLX exposes CLC as intrinsics that wrap the underlying PTX primitive in a structured producer/consumer protocol. A kernel allocates a \emph{CLC pipeline context} via \texttt{tlx.clc\_create\_context}, which materializes per-stage mbarriers tracking \emph{empty} and \emph{full} slots and per-stage 16-byte response buffers in shared memory that the hardware writes asynchronously. The producer side (\texttt{tlx.clc\_producer}) waits for an empty slot, publishes the expected response size via \texttt{barrier\_expect\_bytes}, and issues the request; the consumer side (\texttt{tlx.clc\_consumer}) waits for the full-slot barrier, decodes the returned tile id, and arrives on the empty-slot barrier to release the stage. The tile id returned by \texttt{clc\_consumer} is \texttt{-1} when no work remains, providing a natural termination condition for persistent loops.

\vspace{0.5em}
\noindent \textbf{Distributed Shared Memory.}
Within a cluster, recent NVIDIA architectures expose DSM: a CTA can directly access the shared memory of another CTA through a dedicated shared-cluster address space. TLX adopts DSM as a first-class abstraction because clustered kernels are often producer--consumer \emph{across} CTAs, not just within a CTA, and because paired-CTA collectives often require exchanging tiles or synchronization state without round-tripping to global memory. TLX exposes DSM through explicit storage kinds and remote-access intrinsics. A buffer allocated with \texttt{tlx.storage\_kind.smemCluster} denotes a shared-memory region addressable within the cluster, while standard \texttt{smem} allocations remain CTA-private. Given a local buffer, a CTA can form a \emph{remote view} into another CTA's shared memory via \texttt{tlx.remote\_view()}; TLX also provides direct shared-memory store primitives, with mbarrier completion, to move data to a peer CTA.

\vspace{0.5em}
\noindent \textbf{Multi-CTA Instructions.}
Recent NVIDIA architectures also expose \emph{multi-CTA} instructions, where a small fixed group of CTAs in a cluster participates in a single collective operation. A representative example is Blackwell’s generation-5 tensor cores, which support \emph{paired-CTA MMA}: two CTAs jointly issue one tensor-core instruction, contribute different operand fragments, and cooperatively produce the output tile. Unlike two independent CTAs that later reduce partial results, this is a hardware-level collective, so correctness requires matched issue order, compatible operand layouts, and explicit inter-CTA synchronization.

TLX treats paired-CTA MMA as a first-class cluster instruction, but does not infer grouping or data movement from layout metadata. Instead, grouping is expressed explicitly: the user specifies which CTAs participate in the collective and how operands are partitioned across them. TLX provides primitives to exchange operand fragments through DSM when needed and to synchronize participating CTAs so they reach the collective instruction in a consistent state. This design makes multi-CTA execution a programmable composition of cooperating CTAs, rather than a behavior derived implicitly from data layout.

Another important class is \emph{TMA multicast}, which makes data movement itself a cluster-level collective. A single Tensor Memory Accelerator transfer can populate shared-memory tiles for multiple CTAs in a cluster, reducing redundant global-memory traffic. In GEMM mainloops, this aligns with common reuse patterns, where CTAs may share input tiles across rows or columns. In TLX, multicast behavior is not inferred from layout metadata or compiler analysis. Instead, TLX exposes explicit APIs that allow users to specify the set of target CTAs for each transfer. This design enables flexible communication patterns beyond fixed row- or column-wise reuse, allowing multicast groups to be tailored to the computation. As a result, data movement becomes a programmable primitive, and users can directly control when and how tiles are shared across CTAs, along with necessary synchronization to ensure correctness.

\begin{figure*} \centering
        \includegraphics[width=\linewidth]{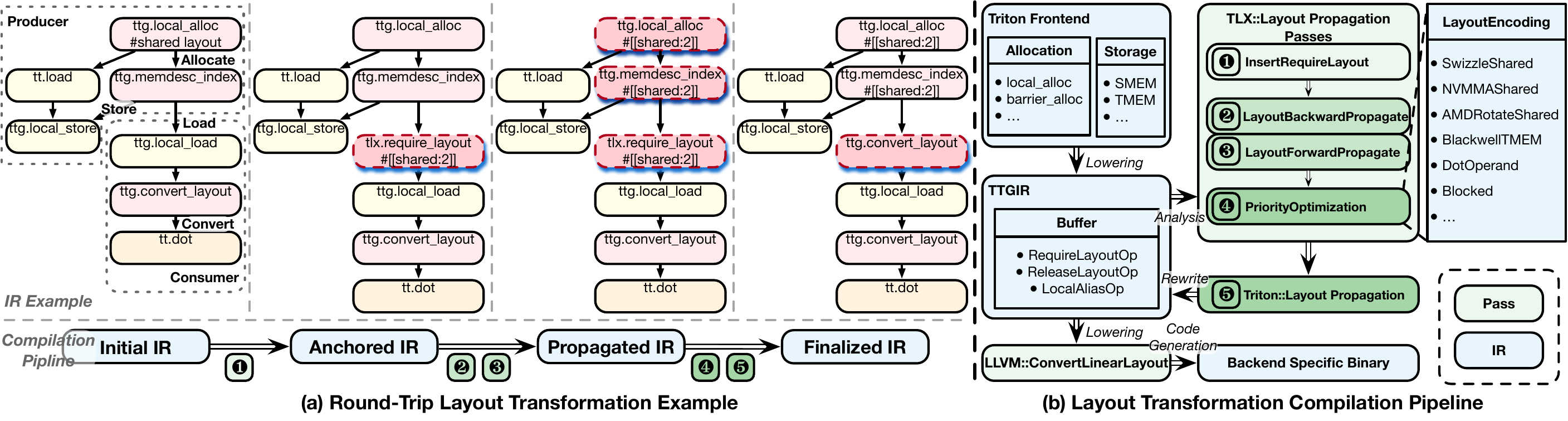}
            \vspace{-2em}
            \caption{Local-memory control and layout propagation in TLX. The left side shows how explicit buffer, alias, and layout operations represent shared local state in the program, while the right side summarizes the compiler flow that propagates, resolves, and lowers those layout constraints.}
            \label{fig:memory-layout}
\end{figure*}

\vspace{0.5em}
\noindent \textbf{Summary.} Cluster programming imposes a rule: a CTA may \emph{arrive} at a remote mbarrier but must not \emph{wait} on one. TLX respects this by keeping producer/consumer slot protocols local to each CTA while expressing inter-CTA coordination using an ``arrive remote, wait local'' pattern when necessary. The compiler inserts cluster-wide synchronization after barrier initialization to prevent races where one CTA signals a remote barrier before the owning CTA initializes it, and it preserves memory ordering by pairing remote stores and arrivals with local waits at the receiving CTA.

The three cluster techniques reinforce each other under TLX's abstraction. CLC provides the dynamic persistent loop that keeps a cluster resident and balanced across irregular tiles; DSM provides the on-chip communication substrate needed for explicit inter-CTA data exchange and producer--consumer coordination; and multicast reduces global traffic by cooperatively staging reused operand tiles into the shared memories of selected CTAs. In TLX, multicast is not inferred from layout but explicitly specified by the user, enabling flexible communication patterns beyond fixed row- or column-wise reuse.

These control mechanisms are based on an explicit shared state. That is the role of TLX's local-memory control: it is the layer that makes warp- and cluster-level coordination precise enough to optimize today and extensible enough to retarget when future hardware changes the available local-memory spaces, layouts, or synchronization rules.

\subsection{Local Memory Control}
\label{subsec:lmm}

Local memory makes the shared state of warp- and cluster-level control explicit. Producer warps allocate and fill staging buffers, consumer warps read them under tight latency constraints, cluster-scoped communication stages reuse those same buffers or their aliases, and later regions may recycle the storage for epilogues or reductions once earlier phases complete. In a uniform single-instruction multi-warp lowering, the compiler can often infer one shared-memory layout because all warps execute the same instructions in the same order. In the MIMW setting, by contrast, different regions and even different CTAs may observe the same logical buffer through different access patterns, and buffers may be intentionally aliased or reused across phases to stay within on-chip capacity. Making these relationships explicit is what ties the first two control layers together in a form that remains analyzable as hardware introduces new memory layouts, new asynchronous transfers, or new collective instructions.

\vspace{0.5em}
\noindent \textbf{Integration with MIMW.}
\texttt{local\_alloc} names the allocation through which tasks communicate, while alias operations describe intentional storage reuse across different logical views of that allocation. Once producers and consumers are decoupled in time, and once cluster-level collectives may read or publish related state asynchronously, the program must make buffer ownership, reuse, and layout requirements explicit to avoid accidental overwrites or mismatched views. At the same time, modern GPUs expose multiple local-memory spaces and hardware operations, each imposing specific layout constraints on the same logical tile. Without compiler-visible abstractions for these behaviors, a MIMW kernel either over-allocates local storage, hurting occupancy, or relies on fragile backend-specific assumptions.

\vspace{0.5em}
\noindent \textbf{Local memory layout propagation.}
\Fig{\ref{fig:memory-layout}} summarizes this local-memory layer, while \Fig{\ref{fig:implementation}} and \Fig{\ref{fig:ws}} show how it fits into the compiler and execution model. TLX introduces explicit buffer and layout constructs while remaining compatible with Triton's native layout machinery. At the Python level, a manually managed buffer is represented as a \texttt{buffered\_tensor} carrying \emph{shape}, \emph{dtype}, \emph{storage kind}, and an optional \emph{layout encoding}. At the IR level, TLX materializes three key TTGIR operations: \texttt{RequireLayoutOp} to enforce a layout constraint, \texttt{ReleaseLayoutOp} to drop a constraint, and \texttt{LocalAliasOp} to represent aliasing between memory descriptors. These operations make layout requirements and reuse decisions explicit where MIMW regions interact.

The propagation algorithm mirrors the pass pipeline shown in the figure. TLX first runs an \emph{insertion} pass that places \texttt{RequireLayoutOp} around values requiring a particular layout, either because the user requested it or because an operation requires it. It then performs \emph{backward propagation} from consumers to producers to infer the layouts that allocations and stores must satisfy, followed by \emph{forward propagation} from producers to consumers to refine how those inferred layouts flow through views, transposes, and warp-specialized region boundaries. Both analyses operate over a dedicated lattice of layout-encoding facts and are aware of TLX control flow, so constraints can traverse region arguments rather than stopping at a syntactic boundary.

After propagation, TLX resolves conflicts and aliasing with a \emph{priority-based optimization}. When multiple constraints meet at a buffer---for example, when a producer prefers a bank-friendly shared-memory swizzle while a consumer requires an MMA-specific encoding---TLX selects a consistent encoding according to a priority policy and rewrites the program accordingly, for example by inserting minimal layout conversions or choosing one canonical encoding for the allocation. When alias groups are present, the optimizer ensures all aliases share a compatible concrete layout and that reuse does not violate constraints across lifetimes. If the constraint system is unsatisfiable, TLX raises a compilation error with explicit diagnostics pointing to conflicting \texttt{RequireLayoutOp} or operation sites.

Finally, TLX converts the resolved encodings into TritonGPU's canonical representation and hands the result to Triton's existing layout propagation and code-generation passes. The result is a single compiler-visible structure in which warp roles, cluster coordination, and local state are all explicit, allowing TLX to absorb new hardware mechanisms by extending these representations rather than by rediscovering whole kernel patterns from scratch.

\section{Implementation}

This section briefly explains how TLX is realized inside Triton's compiler stack. The main implementation goal is to preserve the extra execution and memory structure introduced by TLX---warp-specialized tasks, explicit local state, and cluster-aware control---through Triton's existing multi-level lowering pipeline, rather than flattening them too early.

\begin{figure}
\centering
    \includegraphics[width=\linewidth]{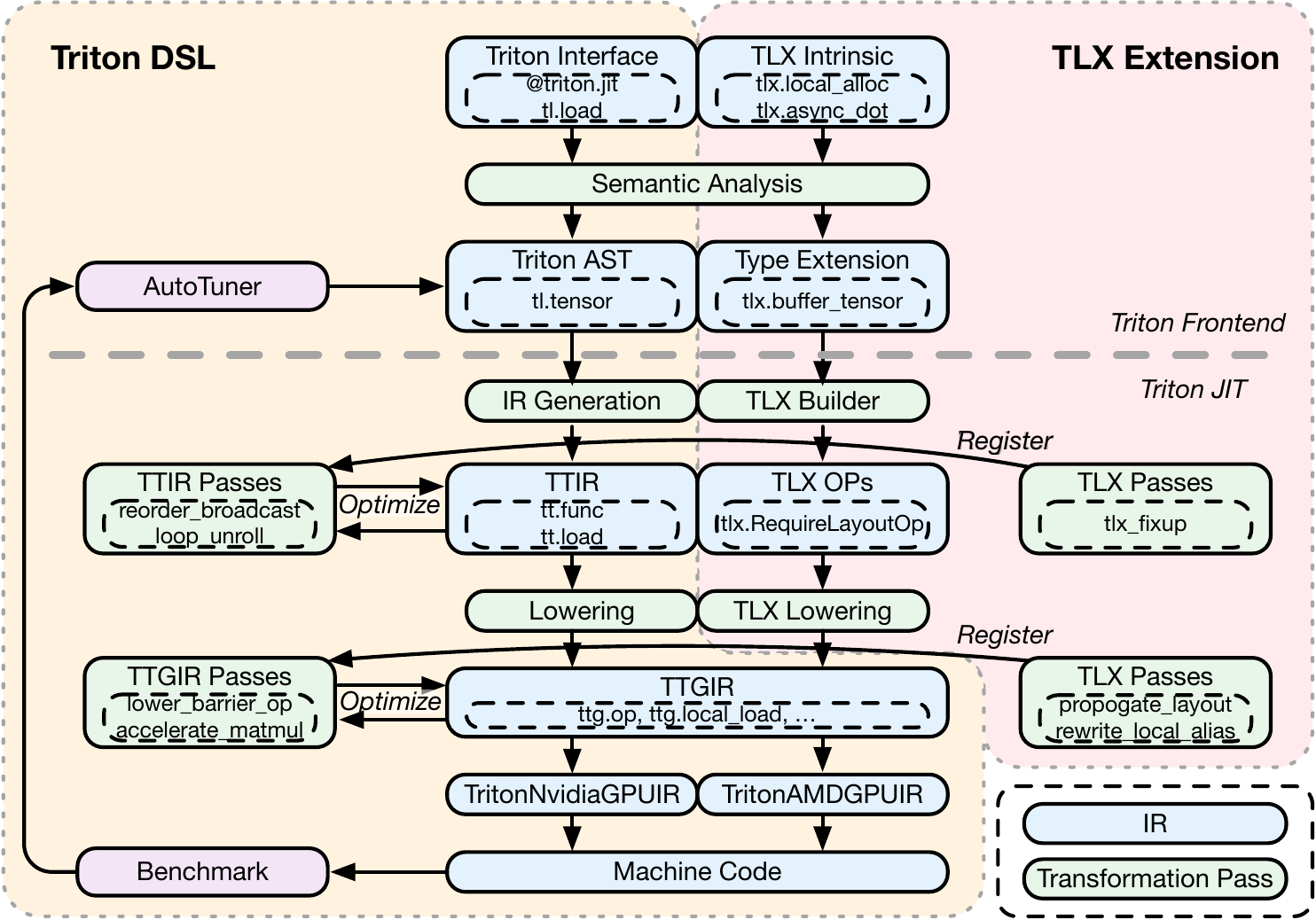}
    \vspace{-2em}
    \caption{TLX's implementation and lowering strategy.}
    \label{fig:implementation}
\end{figure}

The key implementation idea is \emph{additive integration}. TLX is implemented as an extension to Triton's frontend, IR builder, and lowering passes rather than as a separate compiler. This lets TLX reuse Triton's programming model and optimization pipeline, while introducing new compiler-visible constructs only where Triton's original SIMB abstraction is too coarse for MIMW execution. In that sense, this section is about the \emph{systems contribution} of the paper: it shows that MIMW can be integrated into an existing blocked DSL and compiler stack without discarding the parts that already work well. \Fig{\ref{fig:implementation}} summarizes this design.
The implementation has four components. The frontend introduces TLX objects and intrinsics at the source level. IR generation lowers them into explicit TTIR operations. Multi-level lowering preserves those semantics at TTGIR, where target-aware decisions become necessary. Finally, TLX-specific passes resolve layout, synchronization, and cluster-control before backend code generation.

\vspace{0.5em}
\noindent \textbf{Frontend integration.} A TLX kernel remains a standard \texttt{@triton.jit} function, with TLX intrinsics used alongside Triton primitives. During semantic analysis, TLX extends Triton's frontend with the additional entities needed by MIMW programming, especially explicitly allocated local buffers and synchronization objects such as mbarriers. Representing these as first-class frontend objects, rather than opaque intrinsics, makes TLX's execution and memory structure visible to the compiler from the start.

\vspace{0.5em}
\noindent \textbf{IR generation with TLX builder.} After typing, Triton lowers the kernel to TTIR through its standard IR-generation path. TLX augments this step with a builder that maps TLX intrinsics to TLX-specific TTIR operations. Triton's existing TTIR operations continue to represent tile-level tensor computation, while TLX operations represent the additional orchestration semantics, such as task structure, local allocation and views, barrier manipulation, and layout-constrained local loads and stores. This separation is a key systems result: it keeps Triton's tensor computation intact while making MIMW-specific control and memory semantics explicit.

\vspace{0.5em}
\noindent \textbf{Multi-level lowering.} Triton next lowers TTIR to TTGIR, where GPU-specific operations and layout encodings become explicit. TLX extends this conversion so that TLX TTIR operations become TLX TTGIR operations instead of being erased prematurely. This preserves warp-specialized control, local-memory layout requirements, and cluster-scoped communication until the compiler has enough architectural information to lower them correctly. The central implementation requirement is semantic preservation across IR levels.

\vspace{0.5em}
\noindent \textbf{TLX-specific passes and code generation.} TLX adds a small number of passes at TTIR and TTGIR. These passes validate task structure, propagate and resolve local-memory layout constraints, rewrite aliasing and reuse into concrete storage plans, and legalize asynchronous and cluster-aware operations for the target backend. After that, backend lowering maps the resulting TTGIR operations to target instructions such as mbarrier operations, async copy or TMA sequences, and asynchronous tensor-core instructions. The implementation insight here is that TLX does not require a large new backend; it requires preserving high-level orchestration semantics long enough that backend code generation becomes straightforward.


\section{Evaluation}
\label{sec:evaluation}

We evaluate TLX along three dimensions that mirror the logical flow of the paper.
The section is organized around two questions: can TLX match mature implementations on well-studied operators while keeping the code easier to write, and can the same abstractions accelerate deployment to new use cases and optimizations?

\begin{itemize}[leftmargin=*]
    \item Major operators: GEMM and attention\cite{zadouri2026flashattention} show that TLX can reach equivalent performance regimes on the well-studied kernels that dominate frontier-scale ML workloads while using a much more concise programming interface. We benchmark against ATen\cite{pytorch_aten_docs}, PyTorch's\cite{paszke2019pytorch} operator library, because it integrates the strongest practical CUDA and Triton implementations for these operators.

    \item Production use cases: LayerNorm\cite{ba2016layernorm} and multi-GPU GEMM\cite{10.1145/3567955.3567959, jangda2022breaking} show that TLX accelerates development of production-oriented kernels whose optimization depends on orchestration beyond a parallel mainloop, making it easier to deploy new schedules and optimizations.

    \item Backend extensibility: retargeting kernels across NVIDIA and AMD GPUs shows that TLX remains easy to extend across backends as hardware mechanisms evolve.
\end{itemize}

Due to the page limit, we summarize only the key workload regimes here: the appendix covers both canonical and production-skewed GEMM shapes, attention variants across a wide range of sequence lengths, large-$N$ LayerNorm cases that stress reuse and coordination, and multi-GPU GEMM settings that vary both aspect ratio and GPU count. Full per-figure experimental settings are provided in \Sec{\ref{sec:appendix:evaluation_settings}}.

\begin{figure}
    \centering
    \includegraphics[width=\linewidth]{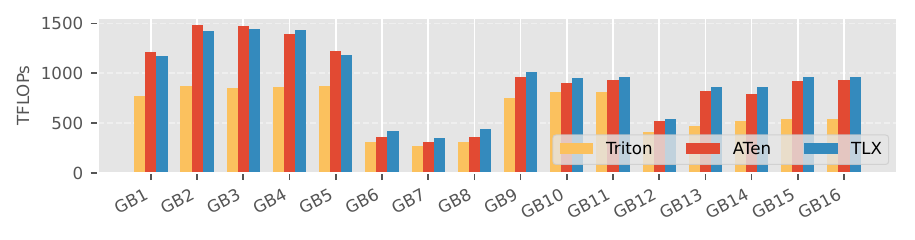}
    \vspace{-3em}
    \caption{GEMM performance on NVIDIA GB200. Detailed settings for each operator are demonstrated in \Sec{\ref{sec:appendix:evaluation_settings}}.}
    \vspace{-1em}
    \label{fig:gemm-performance}
\end{figure}


\subsection{Major Operators}

We begin with the operator families that dominate production ML workloads. Our primary baseline in this subsection is ATen, PyTorch's operator library, because it includes state-of-the-art CUDA and Triton implementations and is therefore the strongest practical comparison point. The goal here is simple: show that TLX remains competitive on mature operators while keeping the source code much closer to Triton than to hand-tuned CUDA.


Production GEMM is not just a tensor-core microbenchmark. Real deployments must handle both balanced matrices and highly asymmetric shapes from projection and feed-forward layers, and they often require backend-specific customization in staging, synchronization, and execution structure. A high-performance GEMM\footnote{\url{https://github.com/facebookexperimental/triton/blob/main/third_party/tlx/tutorials/blackwell_gemm_ws.py}} mainloop in TLX exercises all three technical contributions at once: warp specialization decouples data-staging producers from tensor-core consumers so that TMA loads overlap with MMA compute; local memory management and layout propagation select shared-memory swizzle patterns that avoid bank conflicts across those specialized roles; and cluster launch control keeps the kernel persistent and load-balanced across irregular tile runtimes. \Fig{\ref{fig:gemm-performance}} reports normalized GB200 results on representative production shapes. On these well-studied cases, TLX stays in the same overall performance regime as ATen across both balanced and irregular aspect ratios rather than collapsing on the non-square cases. Importantly, the TLX kernels that achieve this are written in roughly 200 lines of Python-level Triton code, far less than the thousands of lines of CUDA required by a hand-tuned equivalent. The takeaway is that TLX does not trade away performance to recover programmability on a mature operator family.

\begin{figure}
    \centering
    \includegraphics[width=\linewidth]{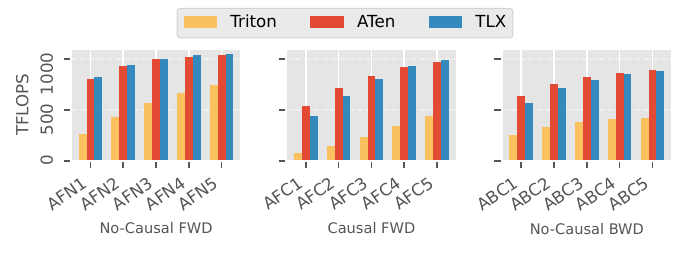}
    \vspace{-2.5em}
    \caption{Attention performance on B200.}
    \label{fig:attention-performance}
\end{figure}

Attention is a stronger test of TLX's customization story because high-performance implementations depend on more than a dense matrix mainloop. A production attention kernel\footnote{\url{https://github.com/facebookexperimental/triton/blob/main/third_party/tlx/tutorials/blackwell_fa_ws_pipelined_persistent.py}} must coordinate tile staging, score computation, masking, normalization, and output accumulation across multiple phases, and the required schedule changes across forward and backward passes as well as causal and non-causal variants. TLX's warp specialization is especially valuable here: different warp groups can own the score-compute, softmax-reduction, and output-accumulation stages, with explicit barrier synchronization connecting each phase rather than forcing all warps through a single uniform loop body. Local memory management then ensures that intermediate tiles such as the attention scores and running softmax statistics are staged on-chip with correct layouts across these phases. \Fig{\ref{fig:attention-performance}} compares TLX against on B200 for no-causal forward, causal forward, and no-causal backward configurations. The results are variant-dependent: TLX trails SOTA results more clearly on the smallest causal-forward cases, but the gap narrows as sequence length grows, and TLX remains close to SOTA results on the no-causal forward and backward configurations over much of the tested range. That is the key point for this subsection: even on another mature and heavily optimized operator family, TLX remains competitive while giving the programmer a cleaner way to express schedule variation than a bespoke CUDA rewrite.

Taken together, the GEMM and attention results establish the first half of the evaluation story: TLX remains in the same performance regime as mature operator implementations while keeping kernels materially easier to write and evolve.

\begin{figure}
    \centering
    \includegraphics[width=\linewidth]{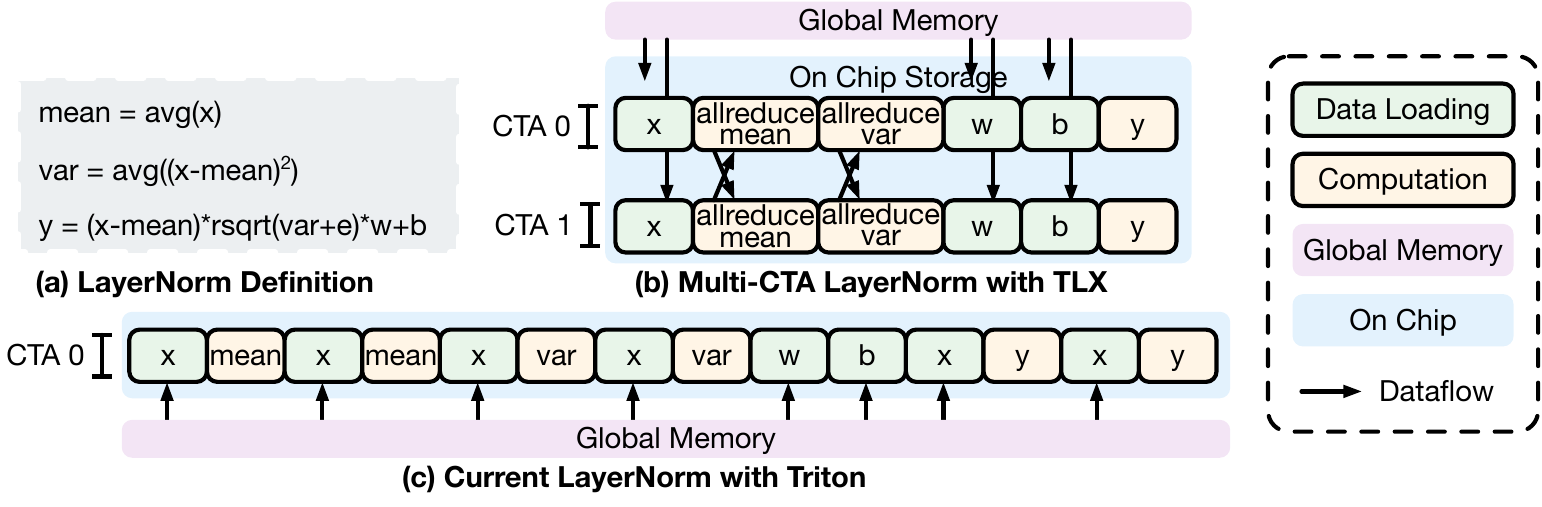}
    \vspace{-2em}
    \caption{Multi-CTA LayerNorm with TLX.}
    \label{fig:layernorm-usecase}
\end{figure}

\begin{figure}
    \centering
    \includegraphics[width=\linewidth]{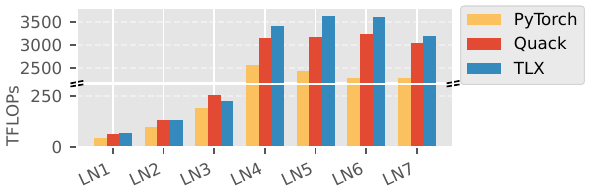}
    \vspace{-2.5em}
    \caption{LayerNorm performance on B200.}
    \vspace{-1em}
    \label{fig:layernorm-performance}
\end{figure}

\subsection{Production Use Cases}

Production systems also contain kernels whose bottlenecks are orchestration rather than raw dense-math throughput. Having established competitiveness on mature operators, we next ask a question: how quickly can the same abstractions be deployed to kernels whose performance depends on new coordination patterns or newly exposed hardware optimizations? Sec.\ref{sec:appendix:use_cases} and the tutorials in our repository contain additional examples, but we focus here on two representative use cases that most clearly expose TLX's strengths beyond dense-kernel code generation: one centered on cluster-local reuse and coordination, and one centered on explicit overlap between communication and compute.

\subsubsection{Multi-CTA LayerNorm}
Large-$N$ LayerNorm is fundamentally a data-movement problem: the same input tile is revisited to compute the mean, the variance, and the final normalized output. \Fig{\ref{fig:layernorm-usecase}} shows how TLX turns that repeated traffic into an explicit cluster-cooperative schedule\footnote{\url{https://github.com/facebookexperimental/triton/blob/main/third_party/tlx/tutorials/blackwell-multi-cta-layernorm_test.py}}. CTAs partition the $N$ dimension, cache their local \texttt{x} tiles on chip, exchange partial sums through DSM, and let one CTA publish the final statistics back to the cluster. The point of the example is not just parallel reduction; it is that TLX makes the reuse path and the synchronization path explicit at the same level as the tile computation.

This is why the example is representative. The same pattern applies whenever large intermediate tiles must survive across phases without paying repeated global-memory traffic, including fused attention and fused GEMM-style kernels. The appendix sketch in \Sec{\ref{sec:appendix:layernorm}} shows the concrete mechanism, but the figure already captures the idea: cluster-level coordination and local reuse are first-class parts of the program. In \Fig{\ref{fig:layernorm-performance}}, Quack~\cite{dao2026quack} denotes a customized implementation built with CuTeDSL. The figure shows that exposing those controls yields practical gains on a bandwidth-bound kernel. More importantly for the argument of this section, the optimization is expressed by extending the same TLX abstractions rather than dropping to a separate low-level implementation path.

\begin{figure}
    \centering
    \includegraphics[width=\linewidth]{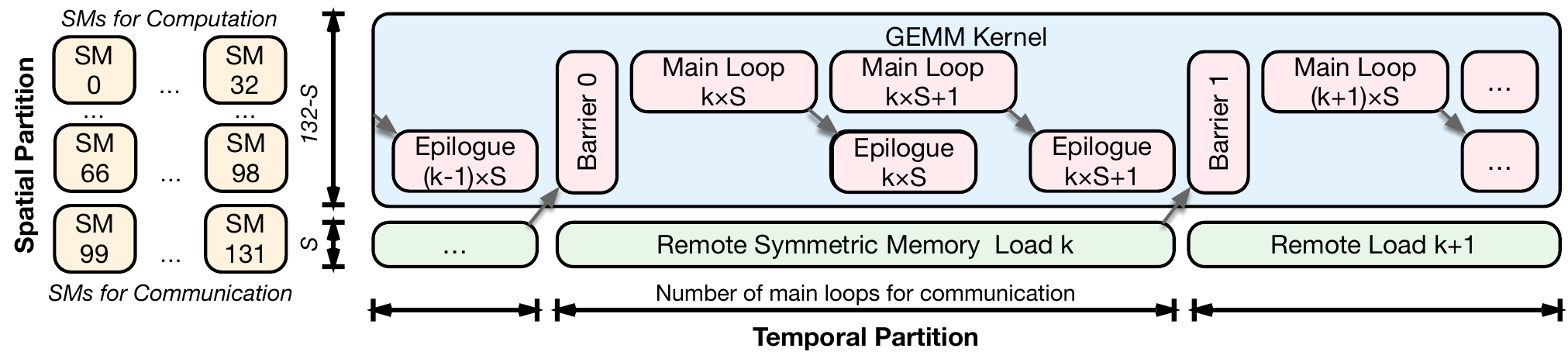}
    \vspace{-2em}
    \caption{Multi-GPU GEMM with TLX.}
    \vspace{-1em}
    \label{fig:dist-gemm-usecase}
\end{figure}

\begin{figure}
    \centering
    \includegraphics[width=\linewidth]{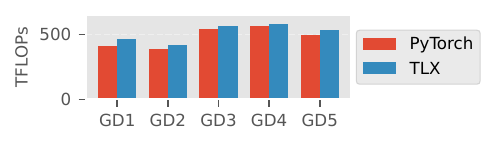}
    \vspace{-2.5em}
    \caption{Performance of multi-GPU GEMM with TLX.}
    \label{fig:dist-gemm-performance}
\end{figure}

\subsubsection{Multi-GPU GEMM}
Multi-GPU GEMM exposes a different orchestration problem: the key question is not how to reuse one tile locally, but how to keep communication and tensor-core compute in flight at the same time. \Fig{\ref{fig:dist-gemm-usecase}} shows TLX's answer. Some CTAs advance the communication stream, others continue the GEMM mainloop, and cluster-visible buffers plus fine-grained barriers connect the two into one pipeline. The important point is that overlap is expressed directly in the kernel structure rather than recovered indirectly from separate launches or coarse global phase boundaries.

This pattern is broader than distributed GEMM. The same role-specialized pipeline appears in fused kernels and megakernel settings, where communication, data staging, and multiple compute stages must coexist inside one larger execution graph. The appendix sketch in \Sec{\ref{sec:appendix:multi_gpu_gemm}} gives the concrete code shape, while \Fig{\ref{fig:dist-gemm-performance}} shows that the resulting schedule is not only expressible but effective in practice. This is the second half of the evaluation story in concrete form: TLX is useful not only for reproducing known fast kernels, but also for deploying new optimizations quickly when the execution structure itself changes.

\begin{figure}
    \centering
    \includegraphics[width=\linewidth]{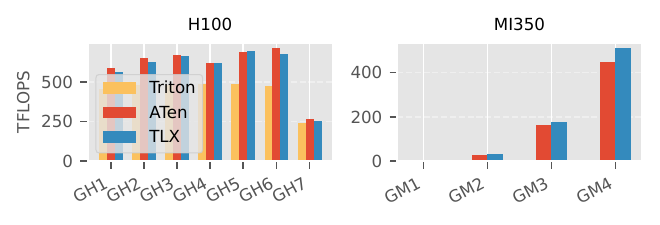}
    \vspace{-2.5em}
    \caption{GEMM results on NVIDIA H100 and AMD MI350. }
    \vspace{-1em}
    \label{fig:gemm-backend-adaptation}
\end{figure}

Taken together, these two use cases show why TLX is effective for fast deployment of new kernel ideas. Its task structure, buffer ownership, and synchronization edges are explicit enough that programmers and agents can reason about them compositionally instead of inferring hidden execution conventions. That clarity matters not only for code generation, but also for adapting a kernel to a new use case, a new overlap pattern, or a new hardware optimization.

\subsection{Backend Extensibility}

Just as importantly, the kernel family must be retargeted as hardware changes. \Fig{\ref{fig:gemm-backend-adaptation}} reports GEMM results on NVIDIA H100 and AMD MI350. On H100, TLX closely tracks ATen across all tested shapes and is nearly indistinguishable from the baseline on several of them. On MI350, TLX matches or slightly exceeds ATen across the problems. This is evidence in the section for backend extensibility: TLX's two-layer design localizes hardware-specific lowering to the extension layer while preserving the same source-level kernel across vendors. The developer writes one warp-specialized, layout-annotated kernel; backend-specific details such as shared-memory swizzle patterns, barrier encodings, and tensor-core instruction selection are resolved during compilation. This completes the evaluation narrative: TLX is competitive on mature kernels, easier to program, and practical to retarget when new hardware or new optimizations appear. In production settings, this matters as much as raw throughput, because hardware evolves faster than any fixed library interface and kernel developers need a path to retarget implementations without rewriting them from scratch.

\section{Conclusion}

TLX introduces a practical MIMW extension to Triton that better matches modern GPU execution. Across major operators, production use cases, and backend retargeting, it shows that explicit orchestration can remain both performant and programmable in a production compiler stack. More broadly, we view TLX as a path toward preserving the productivity of high-level GPU DSLs while exposing the control needed to keep pace with rapidly evolving hardware.


\bibliographystyle{ACM-Reference-Format}
\bibliography{reference} 
\clearpage

\appendix
\section{Evaluation Settings}

\label{sec:appendix:evaluation_settings}

This section summarizes the exact workload settings used by the evaluation figures in the main text. Each table is linked directly to the figure it supports so the reported performance can be interpreted against the corresponding problem family. The goal of these settings is not to exhaustively sweep all possible workloads, but to cover the representative regimes emphasized in the paper: regular dense kernels, production-skewed operator shapes, bandwidth-bound normalization, and communication-overlapped distributed execution.

Table~\ref{tab:b200-gemm-shapes} provides the B200 GEMM settings used for Figure~\ref{fig:gemm-performance}. The first five cases are canonical square GEMMs, which establish a stable reference point for dense tensor-core throughput. The remaining cases intentionally stress asymmetric regimes with large $M$, large $N$, or large $K$, reflecting projection, feed-forward, and expert-style workloads that appear in production training and inference. This mix is important because TLX is meant to remain effective not only on balanced benchmark shapes but also on the irregular aspect ratios that often expose scheduling and staging weaknesses.

\begin{table}[b]
  \centering
  \caption{GEMM shapes on B200}
  \label{tab:b200-gemm-shapes}
  \begin{tabular}{llccc}
    \hline
    ID & Category & M & N & K \\
    \hline
    GB1 & Canonical & 8192 & 8192 & 1024 \\
    GB2 & Canonical & 8192 & 8192 & 2048 \\
    GB3 & Canonical & 8192 & 8192 & 4096 \\
    GB4 & Canonical & 8192 & 8192 & 8192 \\
    GB5 & Canonical & 8192 & 8192 & 16384 \\
    GB6 & Large M & 442368 & 448 & 192 \\
    GB7 & Large M & 589824 & 256 & 128 \\
    GB8 & Large M & 589824 & 448 & 192 \\
    GB9 & Large M & 589824 & 512 & 2048 \\
    GB10 & Large N & 1152 & 32768 & 9216 \\
    GB11 & Large N & 1152 & 32768 & 12800 \\
    GB12 & Large N & 2048 & 64512 & 256 \\
    GB13 & Large K & 512 & 4096 & 64512 \\
    GB14 & Large K & 2304 & 1024 & 32768 \\
    GB15 & Large K & 2304 & 1024 & 63488 \\
    GB16 & Large K & 2304 & 1024 & 65536 \\
    \hline
  \end{tabular}
\end{table}

Table~\ref{tab:h100-gemm-shapes} gives the NVIDIA H100 settings used in Figure~\ref{fig:gemm-backend-adaptation}. These cases retain the same general workload character as the B200 study while using a smaller set of shapes, since the purpose of the figure is backend retargeting rather than an exhaustive platform comparison. The square problems check that TLX preserves strong dense-kernel performance on an earlier NVIDIA architecture, while the irregular cases verify that the same source-level kernel remains competitive when the underlying hardware and lowering strategy change.

\begin{table}[t]
  \centering
  \caption{GEMM shapes on H100}
  \label{tab:h100-gemm-shapes}
  \begin{tabular}{lccc}
    \hline
    ID & M & N & K \\
    \hline
    GH1 & 8192 & 8192 & 1024 \\
    GH2 & 8192 & 8192 & 2048 \\
    GH3 & 8192 & 8192 & 4096 \\
    GH4 & 8192 & 8192 & 8192 \\
    GH5 & 8192 & 8192 & 16384 \\
    GH6 & 2304 & 12800 & 32768 \\
    GH7 & 2285568 & 256 & 256 \\
    \hline
  \end{tabular}
\end{table}

Table~\ref{tab:mi350-gemm-shapes} lists the AMD MI350 settings used in Figure~\ref{fig:gemm-backend-adaptation}. These shapes are intentionally compact and regular because the figure is meant to isolate cross-backend portability rather than reproduce the full NVIDIA-specific workload mix. Using a progression from $256^3$ through $2048^3$ demonstrates that TLX's source-level structure survives backend translation across multiple problem sizes and continues to map well onto a substantially different GPU stack.

\begin{table}[t]
  \centering
  \caption{GEMM shapes on MI350}
  \label{tab:mi350-gemm-shapes}
  \begin{tabular}{lccc}
    \hline
    ID & M & N & K \\
    \hline
    GM1 & 256 & 256 & 256 \\
    GM2 & 512 & 512 & 512 \\
    GM3 & 1024 & 1024 & 1024 \\
    GM4 & 2048 & 2048 & 2048 \\
    \hline
  \end{tabular}
\end{table}

Table~\ref{tab:b200-attention-gemm-shapes} contains the B200 attention settings used for Figure~\ref{fig:attention-performance}. We include causal forward, non-causal forward, and backward configurations because these variants exercise different synchronization and dataflow patterns even when the tensor shapes are similar. The sequence lengths scale from $1024$ to $16384$, which covers the regime where attention kernels transition from latency-sensitive short contexts to bandwidth- and staging-sensitive long contexts. This makes the comparison meaningful for production workloads rather than for a single favorable attention variant.

\begin{table}[t]
  \centering
  \caption{Attention shapes on B200}
  \label{tab:b200-attention-gemm-shapes}
  \resizebox{\linewidth}{!}{
    \begin{tabular}{lccccccccc}
      \hline
      ID & Phase & Causal & M & N & K & P & Q & R \\
      \hline
      AFC1 & FWD & Causal & 4 & 48 & 48 & 1024 & 1024 & 128 \\
      AFC2 & FWD & Causal & 4 & 48 & 48 & 2048 & 2048 & 128 \\
      AFC3 & FWD & Causal & 4 & 48 & 48 & 4096 & 4096 & 128 \\
      AFC4 & FWD & Causal & 4 & 48 & 48 & 8192 & 8192 & 128 \\
      AFC5 & FWD & Causal & 4 & 48 & 48 & 16384 & 16384 & 128 \\
      AFN1 & FWD & Non-Causal & 4 & 48 & 48 & 1024 & 1024 & 128 \\
      AFN2 & FWD & Non-Causal & 4 & 48 & 48 & 2048 & 2048 & 128 \\
      AFN3 & FWD & Non-Causal & 4 & 48 & 48 & 4096 & 4096 & 128 \\
      AFN4 & FWD & Non-Causal & 4 & 48 & 48 & 8192 & 8192 & 128 \\
      AFN5 & FWD & Non-Causal & 4 & 48 & 48 & 16384 & 16384 & 128 \\
      ABC1 & BWD & Non-Causal & 4 & 48 & 48 & 1024 & 1024 & 128 \\
      ABC2 & BWD & Non-Causal & 4 & 48 & 48 & 2048 & 2048 & 128 \\
      ABC3 & BWD & Non-Causal & 4 & 48 & 48 & 4096 & 4096 & 128 \\
      ABC4 & BWD & Non-Causal & 4 & 48 & 48 & 8192 & 8192 & 128 \\
      ABC5 & BWD & Non-Causal & 4 & 48 & 48 & 16384 & 16384 & 128 \\
      \hline
    \end{tabular}
  }
\end{table}

Table~\ref{tab:layernorm-shapes} gives the LayerNorm settings used for Figure~\ref{fig:layernorm-performance}. The chosen cases emphasize large-$N$ normalization, which is exactly the regime discussed around Figure~\ref{fig:layernorm-usecase}: repeated global-memory traffic and reduction coordination become dominant, so cluster-level parallelism and on-chip reuse matter. The small-batch and large-batch settings together show that the optimization is sensible both for latency-oriented shapes and for throughput-oriented shapes that arise in large model execution.

\begin{table}[t]
  \centering
  \caption{LayerNorm evaluation shapes}
  \label{tab:layernorm-shapes}
  \begin{tabular}{lccc}
    \hline
    ID & B & M & N \\
    \hline
    LN1 & 4 & 1 & 16384 \\
    LN2 & 4 & 1 & 32768 \\
    LN3 & 4 & 1 & 65536 \\
    LN4 & 1152 & 1 & 16384 \\
    LN5 & 1152 & 1 & 32768 \\
    LN6 & 1152 & 1 & 65536 \\
    LN7 & 1152 & 1 & 131072 \\
    \hline
  \end{tabular}
\end{table}

Table~\ref{tab:dist-GEMM-shapes} reports the multi-GPU GEMM settings used for Figure~\ref{fig:dist-gemm-performance}. These workloads vary both GPU count and matrix aspect ratio so that the evaluation captures the two pressures highlighted by Figure~\ref{fig:dist-gemm-usecase}: communication volume changes with the distributed partitioning, and overlap opportunities change with the GEMM shape. Including both 2-GPU and 4-GPU cases shows that the pipeline remains meaningful as communication fan-out increases, while the non-square cases demonstrate that the overlap strategy is not restricted to one balanced matrix family.

\begin{table}[t]
  \centering
  \caption{Multi-GPU GEMM evaluation shapes}
  \label{tab:dist-GEMM-shapes}
  \begin{tabular}{lcccc}
    \hline
    ID & \#GPU & M & N & K \\
    \hline
    GD1 & 2 & 8192 & 2048 & 16384 \\
    GD2 & 4 & 8192 & 2048 & 16384 \\
    GD3 & 4 & 8192 & 8192 & 16384 \\
    GD4 & 4 & 4096 & 8192 & 16384 \\
    GD5 & 4 & 16384 & 4096 & 8192 \\
    \hline
  \end{tabular}
\end{table}
\section{Productivity Survey Details}
\label{sec:appendix:productivity_survey}

This appendix summarizes the survey instrument used for the productivity results reported in \Fig{\ref{fig:productivity-survey}}. The survey was conducted with graduate-level students in a GPU programming course, with 127 students involved in total. We use this population because the survey focuses on low-level GPU programming concepts such as thread binding, local-memory control, warp specialization, and cluster-scoped execution, all of which require participants to reason about concrete kernel structure rather than only high-level model APIs.

The questionnaire asked participants to compare four GPU DSLs: Gluon, TLX, TileLang, and ThunderKitten. The introduction shown to students explained that the survey was about GPU programming productivity, that responses would be analyzed only in aggregate, and that the questionnaire should take roughly 10--15 minutes to complete. Rather than asking for abstract impressions alone, the survey presented short code examples for each DSL and asked students to judge how clearly each system exposed the relevant control mechanism.

The instrument was organized into five parts. The first four parts asked students to rate productivity on a 1--5 Likert scale, where 1 meant very low productivity and 5 meant very high productivity. These parts covered: (1) thread binding and memory layout, (2) shared/local memory management, (3) warp specialization, and (4) cluster launch control. The fifth part asked for an overall ranking of the DSLs from 1 to 4. The wording for each section emphasized a specific control surface: for example, the warp-specialization section asked students to consider how directly a DSL lets them assign different warps to producer, consumer, communication, or synchronization roles; the cluster launch control section asked about expressing Blackwell-style cluster coordination and explicitly noted that only TLX and ThunderKitten expose that mechanism directly.

To keep comparisons concrete, each question used short, aligned code excerpts rather than full kernels. The survey therefore measured perceived productivity for expressing representative low-level mechanisms, not only familiarity with one syntax. This design is especially important for TLX, because its contribution is not merely another blocked-kernel DSL interface, but an extension that exposes execution structure usually hidden behind higher-level abstractions. The survey was intended to test whether these additional controls remain understandable and usable in practice for students with graduate-level GPU programming training.

\noindent \textbf{Representative warp-specialization prompt.}
The warp-specialization portion of the survey asked students to compare how the DSLs express producer/consumer partitioning and warp-role assignment. Listing~\ref{lst:appendix:survey_ws_examples} reproduces the representative snippets shown in that part of the questionnaire.

\begin{CodeListing}[Warp-specialization examples used in the survey.]{label=lst:appendix:survey_ws_examples, breakable}{basicstyle=\scriptsize\ttfamily, breaklines}
##### Gluon #####
@gluon.jit
def matmul_load_partition(p, SchedulerImpl: gl.constexpr):
	# Load code here
@gluon.jit
def matmul_mma_partition(p, SchedulerImpl: gl.constexpr):
	# MMA code here
@gluon.jit
def matmul_epilogue_partition(p, SchedulerImpl: gl.constexpr):
	# Epilogue code here
p = PartitionArgs(..., num_warps)
gl.warp_specialize([
	(matmul_epilogue_partition, (p, SchedulerImpl)),
	(matmul_load_partition, (p, SchedulerImpl)),
	(matmul_mma_partition, (p, SchedulerImpl)),
], [1, 1], [24, 24])

##### TLX #####
with tlx.async_tasks():
	with tlx.async_task("default"):
		# Producer (async load)
	with tlx.async_task(num_warps=4, replicate=2, registers=232):
		# Consumers (wgmma + async store)

##### TileLang #####
for ko in T.Pipelined(T.ceildiv(K, block_K), num_stages=2):
	with T.ws(1):
		# Producer
	with T.ws(0):
		# Consumer

##### ThunderKitten #####
if (warpgroup::groupid() < C::NUM_CONSUMERS) {
	// Consumer
	warpgroup::increase_registers<232>();
} else {
	// Producer
	warpgroup::decrease_registers<40>();
}
\end{CodeListing}

The remaining sections of the questionnaire followed the same pattern: each presented short code excerpts for the same conceptual mechanism, then asked students to score how easy the DSL was to understand, adapt, and tune. For thread binding and memory layout, the focus was how kernels map work onto threads, warps, and blocks, and how tensor layouts are expressed. For local-memory management, the focus was allocation, data movement, reuse, and layout control in shared memory or specialized local storage. For cluster launch control, the survey used only TLX and ThunderKitten examples because the other DSLs did not directly expose that feature at the time of the study. Finally, the overall ranking question asked students to aggregate these impressions across all categories.

\section{Use Cases} \label{sec:appendix:use_cases}

This appendix collects compact implementation sketches for the representative use cases discussed in the evaluation. The goal is not to reproduce each kernel in full, but to make the orchestration structure behind the main-text examples concrete enough to inspect. Each subsection therefore focuses on the key TLX mechanism that differentiates the use case from a conventional blocked kernel, while the summary table below provides a quick map of the appendix contents.

\begin{table}[t]
  \centering
  \caption{Summary of appendix use cases}
  \label{tab:appendix-usecase-summary}
  \begin{tabular}{p{0.28\linewidth}p{0.56\linewidth}}
    \hline
    Use Case & Key TLX Mechanism \\
    \hline
    Multi-CTA LayerNorm & Cluster all-reduce, local buffering, DSM-based reuse \\
    Multi-GPU GEMM & Communication--computation overlap with warp-specialized CTA roles \\
    Simplicial Attention & Warp-specialized producer/consumer pipeline for fused attention \\
    \hline
  \end{tabular}
\end{table}

\subsection{Multi-CTA LayerNorm}
\label{sec:appendix:layernorm}

This appendix provides the code corresponding to the multi-CTA LayerNorm use case discussed in \Fig{\ref{fig:layernorm-usecase}} of the main paper. As noted there, \Fig{\ref{fig:layernorm-usecase}} is a demonstrative example intended to highlight the key optimization ideas, rather than to reproduce the implementation line by line. The listings in this appendix give the concrete Triton baseline and the corresponding TLX implementation sketch used to realize that design.

The contrast between the two versions is centered on data movement and reuse. In the Triton baseline, LayerNorm over a large $N$ dimension requires three logical passes over the input: one to compute the mean, one to compute the variance, and one to normalize and apply the affine transform. Because a single CTA cannot retain all intermediate data across these phases, values such as \texttt{x} are loaded repeatedly from global memory. This leads to redundant memory traffic and makes it difficult to overlap communication with computation.

The TLX implementation addresses this issue by distributing the work across a CTA cluster. Each CTA is assigned a slice of the $N$ dimension according to \texttt{tlx.cluster\_cta\_rank()}, computes local partial reductions, and exchanges these partials through distributed shared memory. The helper routine \texttt{cluster\_allreduce} illustrates the main mechanism: each CTA stores its local partial result in cluster-accessible on-chip storage, uses \texttt{tlx.async\_remote\_shmem\_store} to make the value visible to peer CTAs, and then participates in a barrier-coordinated aggregation. In addition, the local tile \texttt{x} is buffered using \texttt{tlx.local\_alloc} and \texttt{tlx.local\_store}, allowing the final normalization step to reuse on-chip data instead of fetching \texttt{x} again from global memory.

This organization captures the three main ideas illustrated in \Fig{\ref{fig:layernorm-usecase}}: multi-CTA parallelism, local on-chip caching, and asynchronous remote communication. A designated aggregator CTA waits locally on the barrier, gathers the cluster-wide partial sums, computes \texttt{mean} and \texttt{rstd}, and publishes the results back to the cluster. The resulting synchronization pattern follows the ``arrive remote, wait local'' style described in the main text: remote arrivals are asynchronous, and only the CTA responsible for aggregation must block. Overall, the TLX version reduces redundant global-memory traffic, improves reuse of loaded tiles, and spreads the reduction work across multiple CTAs.

\begin{CodeListing}[Triton LayerNorm.]{label=lst:appendix:triton_layernorm, float}{basicstyle=\scriptsize\ttfamily}
mean = 0.0
for offset in range(0, N, BN):
  x = tl.load(X+offset+tl.range(0, BN), ...)
  mean += tl.sum(x, axis=0)/N
var = 0.0
for offset in range(0, N, BN):
  x = tl.load(X+offset+tl.range(0, BN), ...)
  var += tl.sum((x-mean)^2, axis=0)/N
rstd = 1.0/tl.sqrt(var + eps)
w = tl.load(W, ...)
b = tl.load(B, ...)
for offset in range(0, N, BN):
  x = tl.load(X+offset+tl.range(0, BN), ...)
  y = (x-mean) * rstd * w + b
  tl.store(Y+offset+tl.range(0, BN), y)
\end{CodeListing}

\begin{CodeListing}[TLX LayerNorm.]{label=lst:appendix:tlx_layernorm, float}{basicstyle=\scriptsize\ttfamily}
rank = tlx.cluster_cta_rank()
bars = tlx.alloc_barriers(2)
tlx.barrier_expect_bytes(...)
tlx.cluster_barrier()

# AllReduce over CTAs in cluster using DSM + barrier
def cluster_allreduce(tile, bar):
  local_buff = tlx.local_alloc([BM,1], dtype, num_ctas)
  tlx.local_store(local_buff[rank], tl.sum(tile, axis=1))
  for peer in tl.static_range(num_ctas):
    if peer != rank:
      tlx.async_remote_shmem_store(...)
  tlx.barrier_wait(bar)
  sum = tl.zeros([BM, 1], dtype)
  for peer in tl.static_range(num_ctas):
    sum += tlx.local_load(local_buff[peer])
  return sum

x = tl.load(X, ...)
x_buff = tlx.local_alloc((BM, BN), dtype, 1)
tlx.local_store(x_buff[0], x)
mean = cluster_allreduce(x, bars[0]) / N
rstd = rsqrt(cluster_allreduce((x-mean)^2, bars[1]) / N + eps)
x = tlx.local_load(x_buff[0])
w = tl.load(W, ...)
b = tl.load(B, ...)
y = (x-mean) * rstd * w + b
\end{CodeListing}
\subsection{Multi-GPU GEMM Overlap}
\label{sec:appendix:multi_gpu_gemm}

This appendix gives a TLX-style sketch for the communication--computation overlap pattern used in the multi-GPU GEMM case study discussed in \Fig{\ref{fig:dist-gemm-usecase}} of the main paper. As in the LayerNorm appendix, the purpose is to make the orchestration structure concrete rather than to present a complete runnable kernel. The listing therefore emphasizes CTA roles, staged buffers, and synchronization edges, while eliding transport-specific pointer arithmetic and tile-indexing details.

The key idea is to partition CTAs inside a cluster into two roles. Communication CTAs advance the all-gather stream, place each arrived operand split into cluster-addressable shared memory, and notify the corresponding compute CTAs by remotely arriving on their local barriers. Compute CTAs wait only on their own \texttt{data\_ready} barriers, consume the staged tiles with \texttt{tlx.async\_dot}, and release ring-buffer slots through \texttt{empty} barriers once the split has been consumed. This follows TLX's ``arrive remote, wait local'' discipline and matches the timeline shown in the main text.

The organization mirrors the three ideas highlighted by \Fig{\ref{fig:dist-gemm-usecase}}. CTA specialization separates communication from tensor-core compute, cluster-visible local buffers make gathered splits reusable by the compute CTAs without an extra global-memory round trip, and the ring-buffer barrier protocol makes overlap explicit. The communication path can be implemented with symmetric-memory gathers, NVSHMEM-style transport, or another backend-specific mechanism; the point of the sketch is that TLX factors this transport from the local orchestration structure.

\begin{CodeListing}[TLX multi-GPU GEMM overlap sketch.]{label=lst:appendix:tlx_multigpu_gemm, breakable}{basicstyle=\scriptsize\ttfamily}
comm_rank = tlx.cluster_cta_rank()
is_comm = comm_rank < NUM_COMM_CTAS

# Ring-buffered cluster storage for gathered operand splits.
a_stage = tlx.local_alloc((BM, BK), dtype, NUM_STAGES,
                          tlx.storage_kind.smemCluster)
b_stage = tlx.local_alloc((BK, BN), dtype, NUM_STAGES,
                          tlx.storage_kind.smemCluster)
empty = tlx.alloc_barriers(NUM_STAGES)
data_ready = tlx.alloc_barriers(NUM_STAGES)

if not is_comm:
  for stage in tl.static_range(NUM_STAGES):
    tlx.barrier_arrive(tlx.local_view(empty, stage))
    tlx.barrier_expect_bytes(
      tlx.local_view(data_ready, stage), 
      ...
    )

tlx.cluster_barrier()

def gather_split_to_stage(stage, split_id, remote_rank):
  remote_a = tlx.remote_view(
    tlx.local_view(a_stage, stage), 
    remote_rank
  )
  remote_b = tlx.remote_view(
    tlx.local_view(b_stage, stage), 
    remote_rank
  )
  gather_remote_a(remote_a, split_id)
  gather_remote_b(remote_b, split_id)

if is_comm:
  for split_id in range(NUM_K_SPLITS):
    stage = split_id % NUM_STAGES
    phase = (split_id // NUM_STAGES) % 2
    tlx.barrier_wait(tlx.local_view(empty, stage), phase=phase)
    for peer in tl.static_range(NUM_COMP_CTAS):
      remote_rank = NUM_COMM_CTAS + peer
      gather_split_to_stage(stage, split_id, remote_rank)
      tlx.barrier_arrive(tlx.local_view(data_ready, stage),
                         1,
                         remote_cta_rank=remote_rank)
else:
  acc = tl.zeros((BM, BN), dtype=tl.float32)
  for split_id in range(NUM_K_SPLITS):
    stage = split_id % NUM_STAGES
    phase = (split_id // NUM_STAGES) % 2
    tlx.barrier_wait(tlx.local_view(data_ready, stage))
    a_tile = tlx.local_load(tlx.local_view(a_stage, stage))
    b_tile = tlx.local_load(tlx.local_view(b_stage, stage))
    acc = tlx.async_dot(a_tile, b_tile, acc)
    acc = tlx.async_dot_wait(0, acc)
    tlx.barrier_arrive(tlx.local_view(empty, stage))

c_desc.store(..., acc)
\end{CodeListing}

\subsection{Simplicial Attention}
\label{sec:appendix:simplicial_attention}

This appendix gives a paper-level TLX sketch for the forward simplicial attention kernel described by Roy et al.~\cite{roy2025fast, clift2019logic} and implemented in the FBGEMM TLX codebase\footnote{\url{https://github.com/pytorch/FBGEMM/tree/main/fbgemm_gpu/experimental/simplicial_attention}}. The goal is the same as in the preceding appendices: make the orchestration structure concrete without reproducing the full implementation line by line. The listing therefore keeps the core execution pattern---warp specialization, staged local buffers, and pipelined overlap between tensor-core GEMMs and softmax updates---while eliding descriptor setup, pointer arithmetic, and edge-case indexing.

The main algorithmic challenge is that 2-simplicial attention is naturally trilinear: each query interacts with one $K_1$ tile and one $K_2$ tile, and the output similarly depends on both $V_1$ and $V_2$. The TLX kernel adopts the tensor-core-friendly reformulation highlighted in the accompanying blog post. Instead of materializing a full three-input primitive, it first forms $Q \odot K_1$ in registers, uses tensor cores for the resulting QK--$K_2$ GEMM, performs the online softmax update, and then applies $V_1$ after the PV--$V_2$ GEMM. This keeps the expensive inner loops in a form that maps naturally to WGMMA while avoiding the prohibitive memory cost of storing fully precomputed intermediate tensors.

The execution structure follows three hardware-aligned ideas. First, the asymmetric sliding window uses a small $W_1$ and a large $W_2$, so all $K_1$/$V_1$ tiles for one query can be loaded once into shared memory while $K_2$/$V_2$ tiles are streamed through a ring buffer. Second, CTA-local warp specialization separates one producer task from one or more consumer tasks: the producer issues asynchronous descriptor loads for $Q$, $K_1$, $V_1$, $K_2$, and $V_2$, while the consumer groups perform the tensor-core math and online normalization. Third, the consumer loop is pipelined so that the current QK GEMM overlaps with the previous PV GEMM, reducing tensor-core bubbles in the steady state. The result is a fused attention kernel in which TLX exposes the same modern GPU techniques used by high-performance FlashAttention-style kernels, but adapted to the more complex 2-simplicial dataflow.

\begin{CodeListing}[TLX 2-simplicial attention forward sketch.]{label=lst:appendix:tlx_simplicial_attention, breakable}{basicstyle=\scriptsize\ttfamily, breaklines}
# Shared-memory staging and producer/consumer barriers.
q_tiles  = tlx.local_alloc((BLOCK_M_SPLIT, HEAD_DIM), dtype, NUM_MMA_GROUPS)
k1_tiles = tlx.local_alloc((1, HEAD_DIM), dtype, W1)
v1_tiles = tlx.local_alloc((1, HEAD_DIM), dtype, W1)
k2_tiles = tlx.local_alloc((BLOCK_SIZE_KV, HEAD_DIM), dtype, NUM_BUFFERS)
v2_tiles = tlx.local_alloc((BLOCK_SIZE_KV, HEAD_DIM), dtype, NUM_BUFFERS)
q_fulls  = tlx.alloc_barriers(NUM_MMA_GROUPS)
k1_full  = tlx.alloc_barriers(1)
v1_full  = tlx.alloc_barriers(1)
k2_empty, k2_full = alloc_stage_barriers(NUM_BUFFERS)
v2_empty, v2_full = alloc_stage_barriers(NUM_BUFFERS)

def producer_stream_kv():
  load_q_tiles(q_tiles, q_fulls)
  load_all_k1_v1_tiles(k1_tiles, v1_tiles, k1_full, v1_full)
  for kv2_idx in range(kv2_start, kv2_end, BLOCK_SIZE_KV):
    stage = stage_id(kv2_idx)
    wait_until_empty(k2_empty, v2_empty, stage)
    async_load_k2_v2(k2_tiles, v2_tiles, stage, kv2_idx)

with tlx.async_tasks():
  with tlx.async_task("producer"):
    producer_stream_kv()

  with tlx.async_task(replicate=NUM_MMA_GROUPS):
    cid = tlx.async_task_replica_id()
    wait_for_q_k1_v1(q_fulls, k1_full, v1_full, cid)

    q_rmem = tlx.local_load(tlx.local_view(q_tiles, cid)) * softmax_scale
    acc = tl.zeros([BLOCK_M_SPLIT, HEAD_DIM], dtype=tl.float32)
    l_i = tl.zeros([BLOCK_M_SPLIT], dtype=tl.float32) + 1.0
    m_i = tl.zeros([BLOCK_M_SPLIT], dtype=tl.float32) - float("inf")

    for kv1_idx in range(num_kv1_trips):
      qk1_rmem = q_rmem * tlx.local_load(tlx.local_view(k1_tiles, kv1_idx))
      v1_rmem = tlx.local_load(tlx.local_view(v1_tiles, kv1_idx)).to(tl.float32)

      for kv2_idx in range(num_kv2_trips):
        stage = stage_id(kv2_idx)
        probs, m_i, l_i, acc = fused_qk_softmax_pv_step(
          qk1_rmem, v1_rmem, acc, m_i, l_i,
          k2_full, k2_empty, k2_tiles,
          v2_full, v2_empty, v2_tiles,
          stage, kv2_idx, ...
        )

    acc = finish_last_pv(probs, v1_rmem, acc, v2_full, v2_empty, v2_tiles)
    desc_o.store(..., (acc / l_i[:, None]).to(tlx.dtype_of(desc_o)))
    tl.store(M_ptr + ..., m_i + tl.log(l_i))
\end{CodeListing}

\end{document}